\documentclass[journal=jctcce, manuscript=article,layout = twocolumn]{achemso}
\setkeys{acs}{articletitle=true}

\usepackage[version=3]{mhchem} 
\usepackage[T1]{fontenc}

\usepackage{booktabs}
\usepackage{color}
\usepackage{graphicx}
\usepackage{dcolumn}
\usepackage{bm}
\usepackage{dcolumn}
\usepackage{gensymb}
\usepackage{siunitx}
\usepackage{etoolbox}
\usepackage{booktabs}
\usepackage{multirow}
\usepackage[table,xcdraw]{xcolor}
\usepackage{tabularx}
\usepackage{makecell}
\usepackage{lipsum}
\usepackage{amsmath}
\usepackage{footnote}

\usepackage{bm}
\usepackage{blkarray}
\usepackage[mathlines]{lineno}

\def\beq{\begin{equation}}
\def\eeq{\end{equation}}
\def\bea{\begin{eqnarray}}
\def\eea{\end{eqnarray}}
\def\brcl{\begin{array}{rcl}}
\def\bccl{\begin{array}{ccl}}
\def\blcl{\begin{array}{lcl}}
\def\err{\end{array}}
\def\fatR{{\bf R}}
\def\fatZ{{\bf Z}}
\def\fatr{{\bf r}}

\def\fatx{{\bf x}}
\def\fatc{{\bf c}}
\def\fatQ{{\bf Q}}

\title{Alchemical normal modes unify chemical space}

\author{Stijn Fias}
\email{fiass@mcmaster.ca}
\affiliation{General Chemistry (ALGC), Vrije Universiteit Brussel (Free University Brussels - VUB), Pleinlaan 2, 1050 Brussel, Belgium}
\affiliation{Department of Chemistry \& Chemical Biology, McMaster University, Hamilton, ON, Canada L8S 4L8}
\author{K. Y. Samuel Chang}
\affiliation{Institute of Physical Chemistry and National Center for Computational Design and Discovery of Novel Materials (MARVEL), Department of Chemistry, University of Basel, 4056 Basel, Switzerland}
\author{O. Anatole von Lilienfeld}
\email{anatole.vonlilienfeld@unibas.ch}
\affiliation{Institute of Physical Chemistry and National Center for Computational Design and Discovery of Novel Materials (MARVEL), Department of Chemistry, University of Basel, 4056 Basel, Switzerland}

\date{\today}

\begin{document}
\begin{abstract}
{\em In silico} design of new molecules and materials with desirable quantum properties  
by high-throughput screening is a major challenge due to the high dimensionality of chemical space.
To facilitate its navigation,
we present a unification of coordinate and composition space in terms of
alchemical normal modes (ANMs) which result from  second order perturbation theory.
ANMs assume a predominantly smooth nature of chemical space and form a basis
in which new compounds can be expanded and identified.
We showcase the use of ANMs for the energetics of the iso-electronic series of
diatomics with 14 electrons,
BN doped benzene derivatives (C$_{6-2x}$(BN)$_{x}$H$_6$ with $x = 0, 1, 2, 3$), 
predictions for over 1.8 million BN doped coronene derivatives,
and genetic energy optimizations in the entire BN doped coronene space. 
Using Ge lattice scans as  reference,
the applicability ANMs across the periodic table is demonstrated for 
III-V and IV-IV-semiconductors Si, Sn, SiGe, SnGe, SiSn, as well as AlP, AlAs, AlSb, GaP, 
GaAs, GaSb, InP, InAs, and InSb.
Analysis of our results indicates simple qualitative structure property
rules for estimating energetic rankings among isomers. 
Useful quantitative estimates can also be obtained when few atoms are changed to
neighboring or lower lying elements in the periodic table.
The quality of the predictions often increases with the symmetry of
system chosen as reference due to cancellation of odd order terms.
Rooted in perturbation theory the ANM approach promises to generally
enable unbiased compound exploration campaigns at reduced computational cost.

\end{abstract}

\maketitle

\section{Introduction}
A quantum mechanics based understanding of chemical compound space (CCS) is crucial
for gauging the predictive power and versatility of theoretical chemistry models, as well as
and for the computational design of molecular and solid matter.
Due to its universality to account for the physics of electrons which govern the behavior of matter
the use of quantum mechanics is mandatory in this context.
The complexity of its solutions, however, hampers the intuitive understanding 
and conceptualization of the solutions obtained.
High-throughput-screening campaigns have therefore been 
proposed to tackle materials design challenges~\cite{ComputationalMaterialsDesign_MRS2006,Jain2011,Curtarolo2013},
and extensive materials quantum data records have been established~\cite{MaterialsProject,OQMD,DataPaper2014,smith2017ani}. 
Still, the high dimensionality of CCS~\cite{ChemicalSpace,mullard2017drug} 
combined with the considerable cost for repeatedly evaluating quantum properties from scrach
severely hampers even the most sophisticated optimization algorithms, let alone screening.

\begin{figure}
\centering
\includegraphics[width=8.5cm]{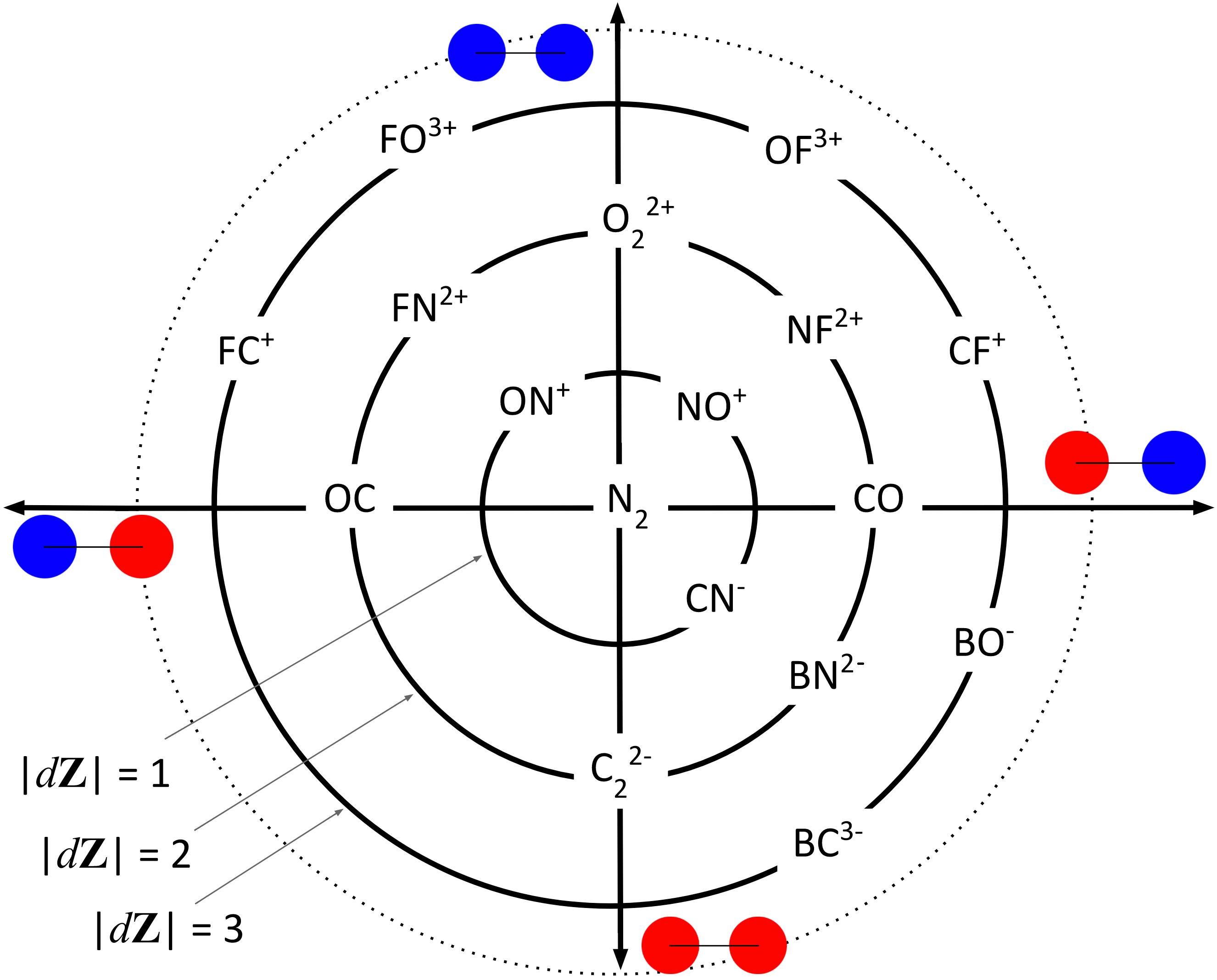}
\caption{
The shell-structure of alchemical hyper-spheres, illustrated for the chemical space of all diatomics with 14 electrons (ANM$_1$ = $1/\sqrt{2}-1/\sqrt{2}$, ANM$_2$ = $1/\sqrt{2}+1/\sqrt{2}$, $\epsilon_1 = -3.65$ a.u., $\epsilon_2 = -2.987$ a.u..
Homo-nuclear diatomics (vertical axis) correspond to the ridge. Interatomic distance dependence is shown in \ref{fig:N2} for all neutral diatomics (horizontal axis).
}
\label{fig:AHS}
\end{figure}


Given thousands of previously acquired representative reference examples used for training, 
quantum machine learning models have recently emerged as a viable
option to further accelerate materials design by multiple orders of
magnitude~\cite{RuppPRL2012,QMLessayAnatole,ML4Polymers_Rampi2013,ML4Crystals_Wolverton2014,GrossMLCrystals2014,Elpasolite_2016},
reaching prediction errors on par with DFT~\cite{googlePaper2017,FCHL}.
Alas, also these methods require representative training sets, and 
the combinatorial nature of chemistry simply prohibits the establishment
of a comprehensive encyclopedia. Consequently, more powerful approaches are needed,
e.g.~exploiting a more rigorous notion of chemical space~\cite{anatole-ijqc2013}.
Here, we investigate such an alternative, physics- rather than statistics-based approach for the sampling of CCS 
which reaches machine learning speed and accuracy. Instead of massive training sets which must be representative,
it requires only a single reference calculation which must be relevant.
It is rooted in second-order perturbation theory and 
includes variations in nuclear charges, a.k.a. ``alchemical changes''.
Alchemical perturbations have been used in quantum chemistry every since H\"uckel's
work on predicting substituent effects in benzene~\cite{HuckelBenzol1931}, 
and Pauling's follow up work~\cite{PaulingBenzol1935,CoulsonBenzol1947}.
More modern approaches include Refs.~\cite{WilsonsDFT,PolitzerParr,ConcavityMezey1985,Marzari_prl_1994,Geerlings_jctc_2000,anatole-prl2005,RCD_Yang2006,ArianaAlchemy2006,anatole-jcp2009-2,CatalystSheppard2010}, and 
more recently, substantial progress has been made along similar lines~\cite{LesiukHigherOrderAlchemy2012,CCSexploration_balawender2013,Samuel-CHIMIA2014,Samuel-JCP2016,AlchemyAlisa_2016,MoritzBaben-JCP2016,Yasmine-JCP2017,saravanan2017alchemical,StijnPNAS2017,balawender2018exploring} 
using first and second order perturbations.  
Here, we use second order perturbation theory to introduce alchemical normal 
modes (ANMs), resulting from diagonalization of a unified Hessian,
to form a complete, low-dimensional, and intuitive basis which spans CCS.
Building on this, we provide a novel understanding of the structure of chemical space, and we show how to 
utilize it for solving inverse design problems with unprecedented speed and accuracy.
The expansion of individual query molecules in their ANMs 
enables rapid energy estimates which we demonstrate for screening over 1.8 M 
BN-doped coronene derivatives based on a single quantum reference calculation.

The remainder of this paper is structured as follows, 
we introduce the theoretical underpinnings of the ANMs in the Theory section, 
exemplifying their usage for molecular nitrogen which is sufficiently 
simple to easily gain an intuition. 
Subsequently, we demonstrate and assess the performance of ANMs for the
complete CCS of all BN doped benzene derivatives.
ANMs of coronene are then used to (i) predict electronic energies of over 1.8 M of its BN doped derivatives,
and (ii) to discover those structures with lowest and highest lying energy, as identified by a genetic optimization algorithm. 
The applicability of ANMs is also demonstrated for solid systems, as exemplified for 
III-V and IV semi-conductors. 
After the discussion of our results we briefly conclude this investigation. 
Finally, methodological details are given for the computational aspects. 

\section{Theory}
Within the Born-Oppenheimer approximation, the total potential ground-state energy of a compound, $U = E + V_{NN}$, 
consists of the nuclear Coulomb repulsion ($V_{NN} = \sum_{I>J} Z_I Z_J/|\fatR_I - \fatR_J|$) and
the electronic energy $E$, the solution of the electronic Schr\"odinger equation (SE). 
In order to facilitate the discussion, all results and discussions in the following
will be concerned exclusively with the latter. Subsequent addition of the $V_{NN}$-term, 
often necessary when aiming for comparison to experimental numbers, is trivial since
composition and coordinates are always assumed to be known.
From the quantum mechanical point of view of the potential energy hyper-surface, 
systems differ only by nuclear charges $\{Z_I\}$, 
atomic coordinates $\{{\bf R}_I\}$, and number of electrons $N$.
Within second order perturbation theory, 
we can therefore Taylor expand the electronic energy of any target system $\fatx^t$
around the electronic energy of a reference system $\fatx_0$,
\bea
E(\fatx^t) & = & E(\fatx_0) + {\bf g} d{\bf x} + \frac{1}{2} d\fatx^{\rm T} {\bf H} d\fatx + \cdots 
\eea
where $\fatx = (Z_1, Z_2, \cdots, Z_M, {\bf R}_1, {\bf R}_2, \cdots, {\bf R}_M, N)$, and
${\bf g}$ and ${\bf H}$ represent a unified gradient and Hessian, respectively. 
First order terms are firmly established for all variables through 
the Hellmann-Feynman theorem for changes in nuclear 
positions (to relax or run ab initio molecular dynamics~\cite{tuckerman_book_SM}),
and charges~\cite{WilsonsDFT,anatole-prl2005,anatole-jcp2006-2,anatole-jcp2007,anatole-jctc2007,LesiukHigherOrderAlchemy2012,CCSexploration_balawender2013,AlchemyAlisa_2016,MoritzBaben-JCP2016}.
The derivative with respect to $N$ is related to ionization potential
and electron affinity by virtue of Koopman's and Janak's theorem~\cite{Janak}, and
exhibits the well established derivative discontinuity at integer $N$~\cite{DD_Perdew1,DD_Perdew2}, 
so important for the construction of improved exchange-correlation approximations~\cite{DiscontinuousXC_MoriSanchezCohenYang2009}.

\begin{figure}
\centering
\includegraphics[width=9cm]{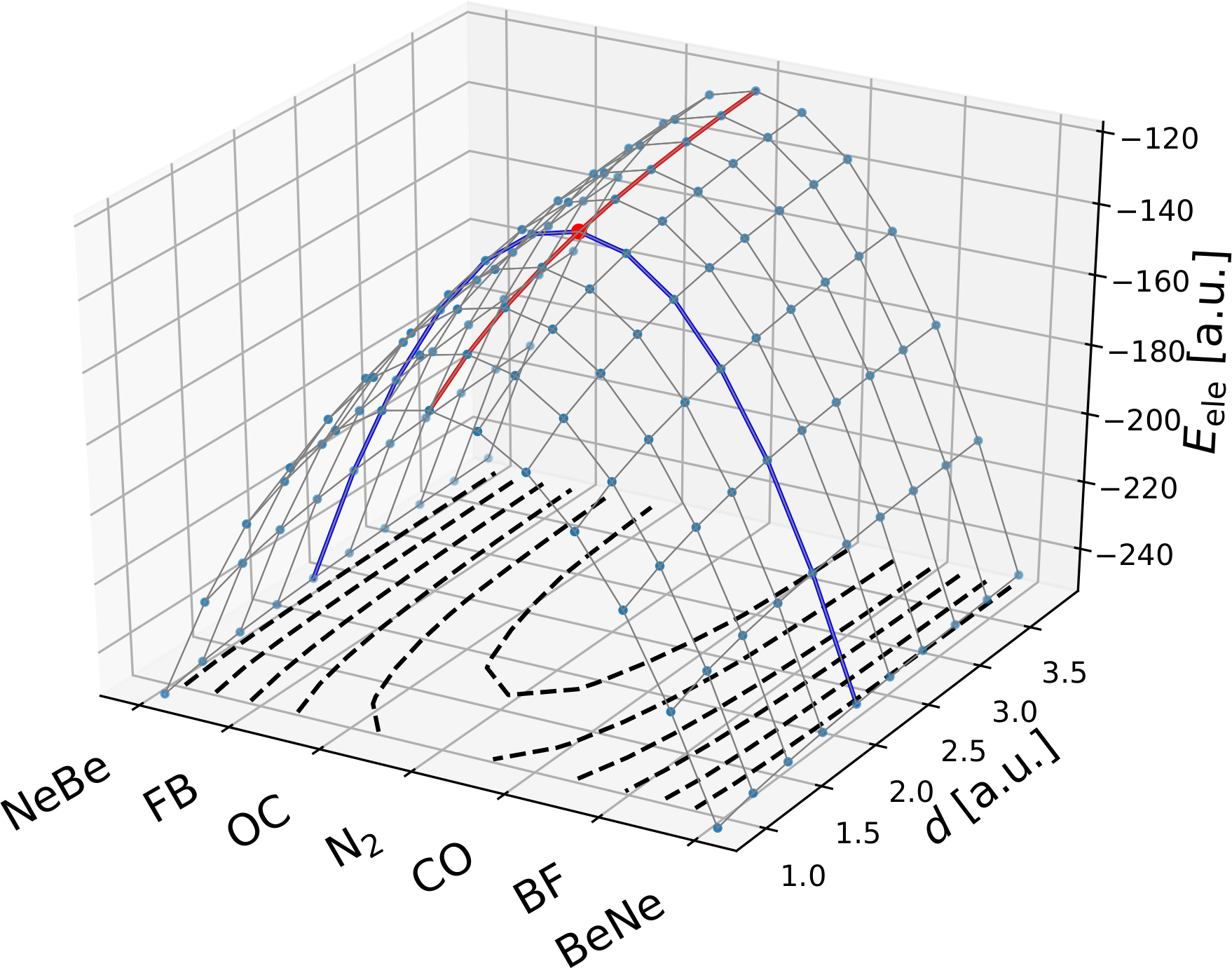}
\includegraphics[width=8.5cm]{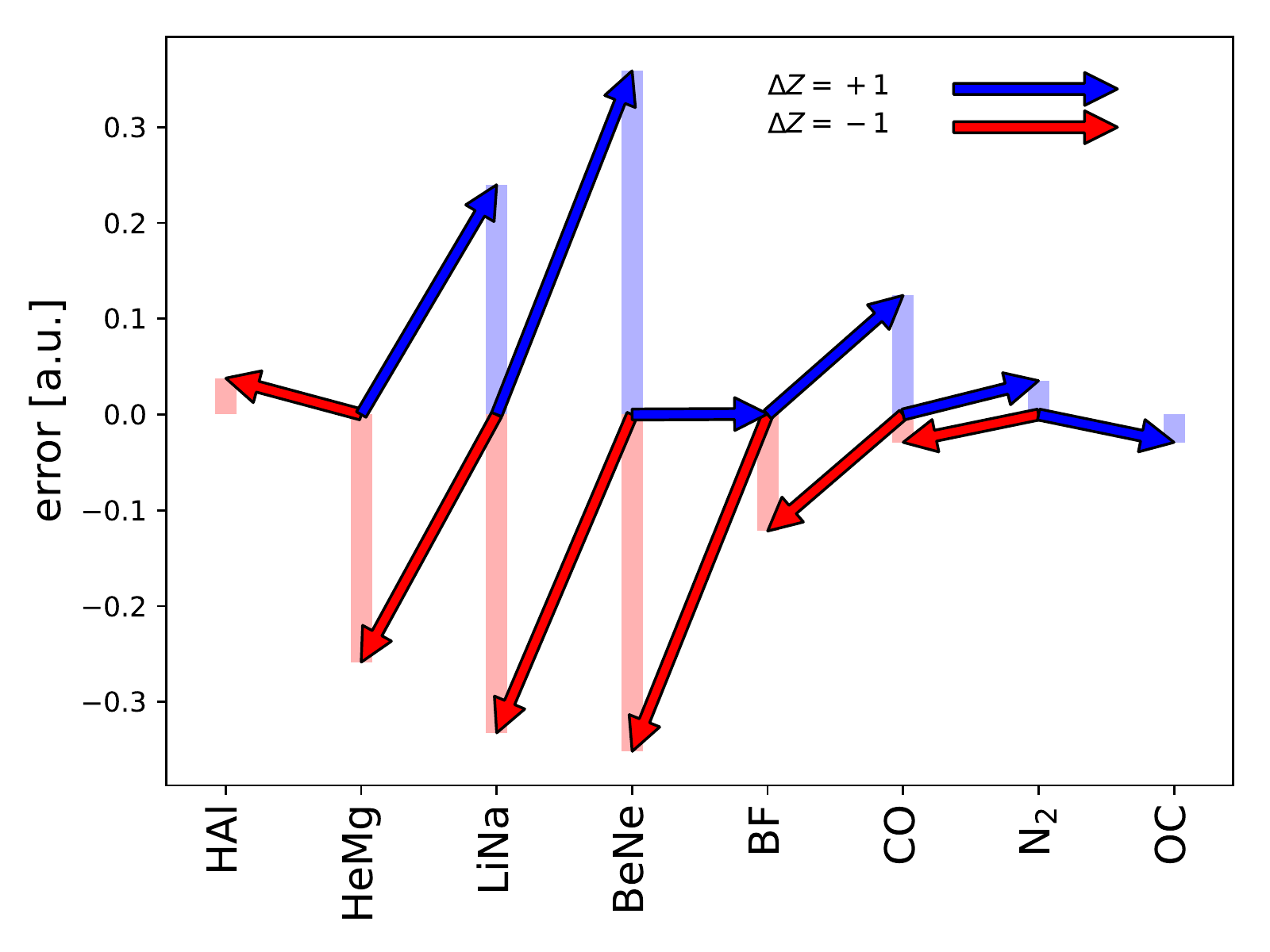}
\caption{
TOP: The electronic energy is shown as a function of interatomic distance and difference in nuclear charge. 
BOTTOM: The error of ANM based predictions of neighboring systems is shown
at fixed interatomic distance at 1.1 {\AA}. 
}
\label{fig:N2}
\end{figure}

Some elements in the Hessian, 
\bea 
{\bf H} = 
\begin{bmatrix}
\\
\frac{\partial^2 E_0}{\partial Z_I \partial Z_J} & \frac{\partial^2 E_0}{\partial Z_I \partial {\bf R}_J} & \frac{\partial^2 E_0}{\partial Z_I \partial N} \\
\\
\frac{\partial^2 E_0}{\partial {\bf R}_I \partial Z_J} & \frac{\partial^2 E_0}{\partial {\bf R}_I \partial {\bf R}_J} & \frac{\partial^2 E_0}{\partial {\bf R}_I \partial N} \\
\\
\frac{\partial^2 E_0}{\partial N \partial Z_J} & \frac{\partial^2 E_0}{\partial N \partial {\bf R}_J} & \frac{\partial^2 E_0}{\partial N^2} \\
\\
\end{bmatrix}
\label{eq:full_hessian}
\eea
are also part of text-book chemistry: 
The coordinate subspace matrix corresponds to the conventional Hessian, related to the harmonic molecular vibrational normal modes, or the second order derivative of the electronic energy with respect to the number 
of electrons is the chemical hardness, introduced by Parr and Pearson~\cite{parryang}.
The $\frac{\partial^2 E_0}{\partial Z_I \partial Z_J}$ block corresponds to 
the alchemical hardness~\cite{anatole-jcp2006-2,Samuel-JCP2016}.
The least conventional off-diagonal blocks correspond to nuclear Fukui functions,
$\frac{\partial^2 E_0}{\partial N \partial {\bf R}_J}$~\cite{baekelandt1996nuclear},
alchemical Fukui Functions $\frac{\partial^2 E_0}{\partial N \partial Z_J}$~\cite{anatole-jcp2006-2,anatole-jcp2007},
and the alchemical force, 
$\frac{\partial^2 E_0}{\partial {\bf R}_I \partial Z_J} = \int d\fatr (\partial_{{\bf R}_I}\rho(\fatr))/|\fatr-\fatR_J| = \int d\fatr (\rho(\fatr) + Z_J \partial_{{\bf Z}_J}\rho(\fatr))(\fatr - \fatR_I)/|\fatr-\fatR_I|^3$ (due to Maxwell-relation).
To the best of our knowledge, such a unified Hessian has not yet been studied in full,
despite the well-known non-linearities of quantum properties in chemical space.

\begin{figure*}
\centering
\includegraphics[width=16cm]{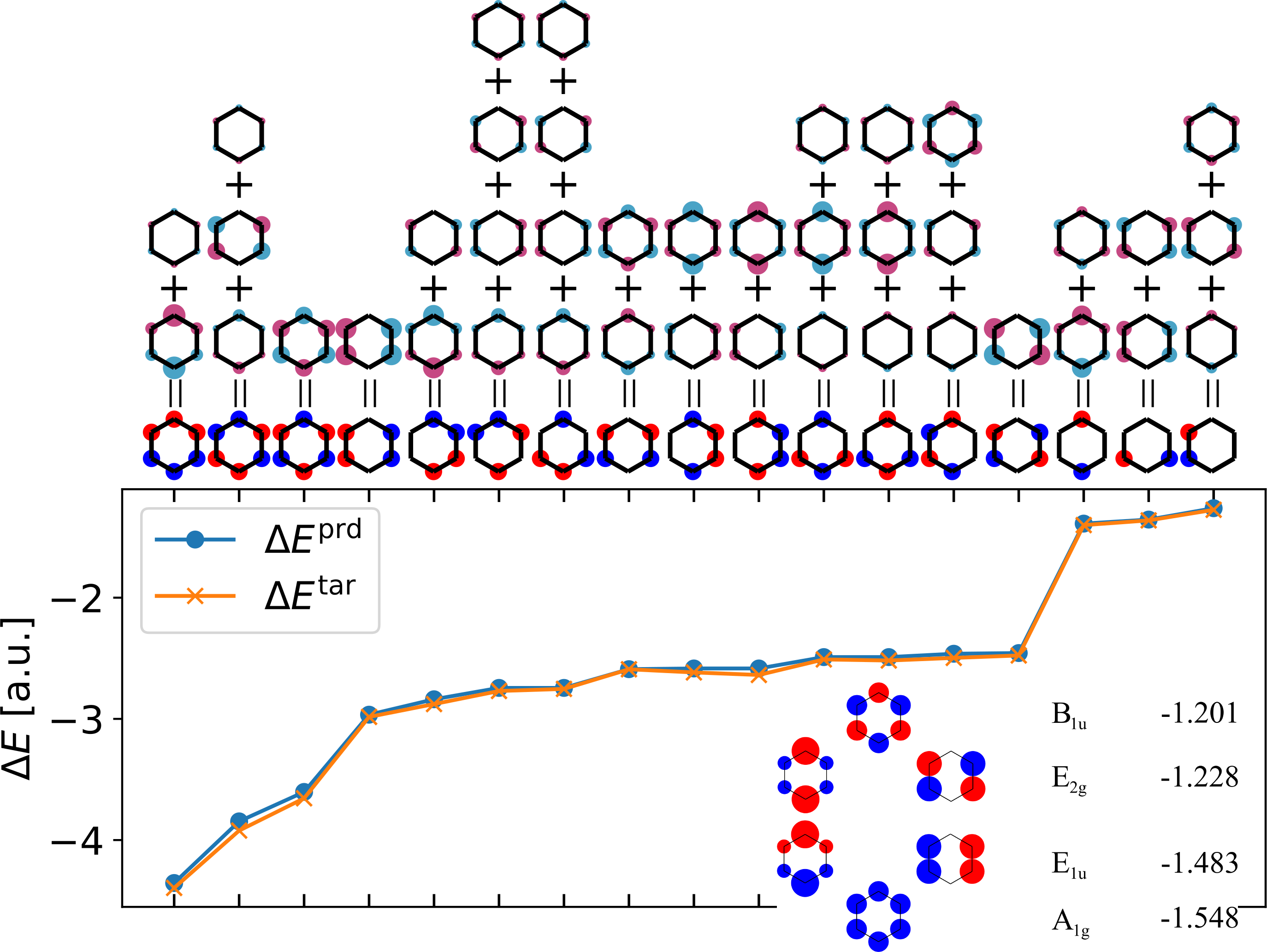}
\caption{
TOP: Expansion of BN doped benzene mutants in alchemical normal modes of benzene (ordered by eigenvalue (Ha)). 
BOTTOM: ANM based predicted electronic energy and corresponding target energy of each mutant in ascending order.
ANMs of benzene and eigenvalues are shown as inset.
}
\label{fig:benzeneAll}
\end{figure*}

In analogy to vibrational normal modes, diagonalization of this unified Hessian 
for any meaningful reference system defines an orthogonal and complete basis in 
which other chemical compounds and their intra-molecular motion can be expanded.
The resulting eigenvalues and eigenfunctions, the ``alchemical normal modes'' (ANMs), 
correspond to principal curvatures, and thus carry fundamental importance for 
our understanding of CCS (Transformation from Hessian matrix to second fundamental 
form may be required).
The composition of any target molecule 
can be linearly expanded in the complete vector basis 
spanned by the ANM matrix ($\fatQ$) of the 
reference compound $\fatx_0$, i.e.~$\fatx^t = \fatx_0 + d\fatx$.
The necessary coefficients are given by $\fatc = \fatQ d\fatx$, resulting in the
second order energy estimate, 
\bea 
E(\fatx^t) & \approx & E(\fatx_0) + {\bf g} d{\bf x} + \frac{1}{2} \fatc^{\rm T} \mathbf{\Upsilon} \fatc 
\label{eq:Predict}
\eea 
where $\mathbf{\Upsilon}$ is the diagonal eigenvalue matrix of the unified Hessian  $\mathbf{H}$.
This framework leads to an encompassing definition of the structure of CCS which
couples configurational, compositional, and electronic degrees of freedom.
Let us consider projections onto lower dimensional manifolds of this structure.
Firstly, for fixed composition ($|_{\{Z_I\}}$), 
the conventional picture of changes in configurations (geometry)
and electron number (redox-properties), emerges.
Secondly, when fixing geometry and electron number for $n$ atom systems instead, 
an $n$-dimensional alchemical hyper-sphere (AHS) can be defined for 
reference compounds with maximal symmetry ({\em vide infra} why) 
being at the origin (i.e.~that system for which all atoms have same nuclear charge, 
$Z_I = N_p/n\;\forall \; I$ where $N_p = |\fatZ|$). 
The AHS has a shell structure where integer nuclear charge combinations
emerge for integer radii, $|d\fatZ|$, i.e.~systems with a correspondence in reality. 
Fig.~\ref{fig:AHS} illustrates the AHS for the di-atomics 
with $N =$ 14 electrons which can be expanded in ANMs of molecular nitrogen. 
ANM $q_1$ corresponds to charge-neutral simultaneous 
depletion and growth of the nuclear charge at the two respective
atomic sites, covering the series N$_2$, CO, BF, ..., AlH, Si.
ANM $q_2$ corresponds to the simultaneous addition or removal
of protons at the two respective atomic sites, covering the series
..., B$_2^{4-}$, C$_2^{2-}$, N$_2$, O$_2^{2+}$, F$_2^{4+}$, ...
Linear combinations of $q_1$ and $q_2$ define all the other possible diatomics
which can be defined on shells with radii $|d\fatZ| = 1, 2, 3, ...$. 
e.g.~NO$^+$, expanded in ANMs of N$_2$, corresponds to $1/\sqrt{2} q_1 + 1/\sqrt{2} q_2$. 
Obviously, while target compounds with large $q_2$ component 
will be increasingly charged and unstable without changes in electron number, 
in the absence of external fields or extreme conditions, 
this extended unified structure of CCS is general in scope 
as it accounts for a continuum of ``alchemical'' chemistries with fractional nuclear charges.
We note that extensions of reality to include such fictitious 
degrees of freedom have a long-standing track-record in thermodynamics and statistical
mechanics, e.g.~in the form of extendended Lagrangians, and can be used 
for any state function.

\begin{figure}
\centering
\includegraphics[width=8.5cm]{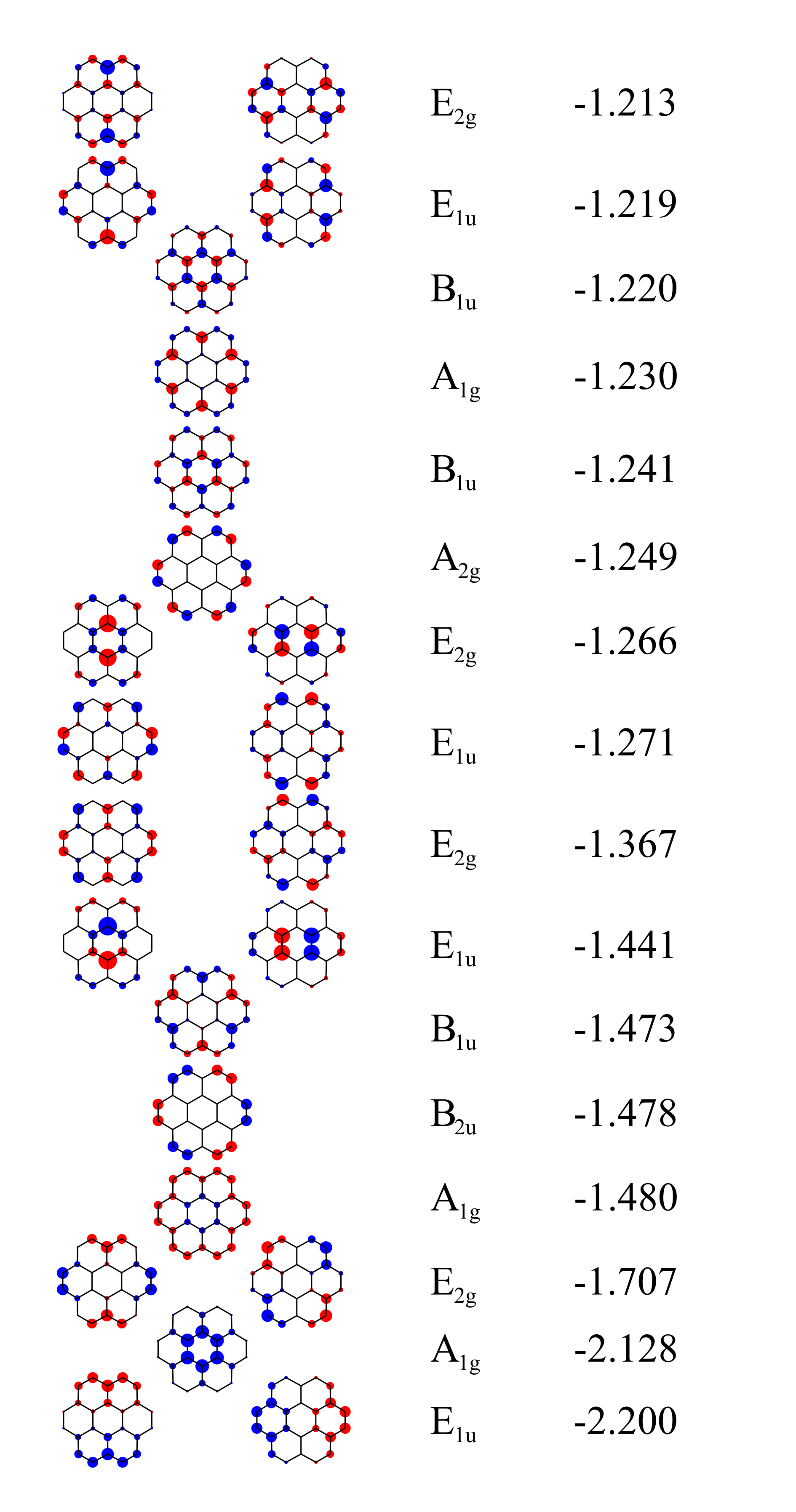}
\includegraphics[width=8.5cm]{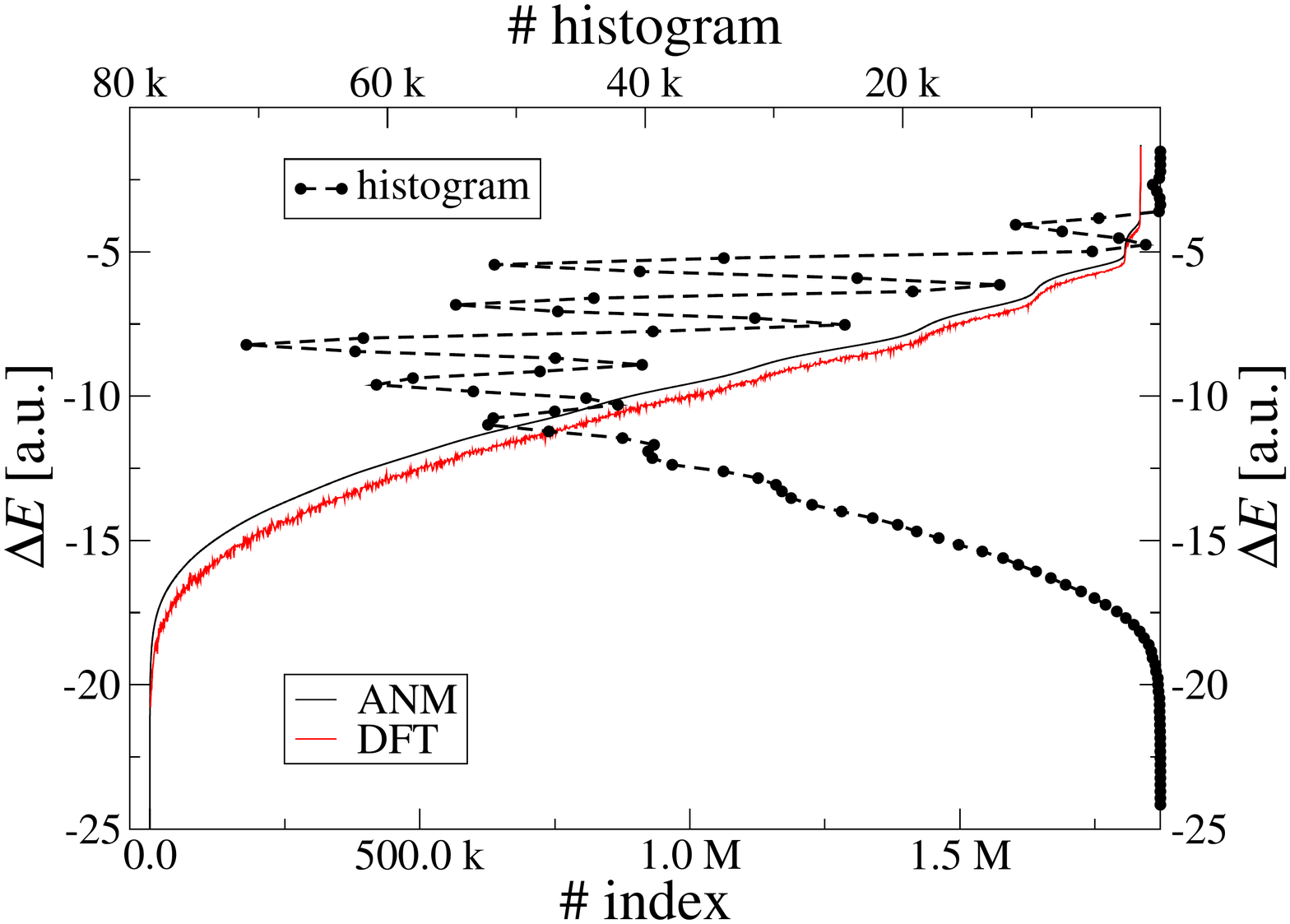}
\caption{
TOP: Alchemical normal modes of coronene and corresponding eigenvalues in Hartree.
BOTTOM: ANM based estimates of energy change from coronene for
$\sim$1.8 M BN doped coronene mutants in ascending order (black solid). 
Their distribution is shown as a histogram (black dashed).
Validating DFT results for sub-sample of $\sim$2 k examples  
shown for comparison (red).
}
\label{fig:Diagram2}
\end{figure}

\section{Results and discussion}
Within the first subsection we present and discuss results obtained for the neutral iso-electronic
diatomic series with 14 electrons, including all interatomic distances and all possible nuclear
charge combinations. 
In order to facilitate the discussion and visualization of results in the subsequent subsections, 
we restrict ourselves to fixed geometries and electron number, and we
focus on changes in composition only without any loss of generality. 
This restriction is obviously severe for large, high-dimensional systems which sample
many effective degrees of freedom, e.g.~proteins with many shallow conformational minima.
However, for materials classes with rigid lattices and an effectively low dimensionality
e.g.~crystals with high symmetry under ambient conditions, 
the relevant configurational degrees of freedom can easily be scanned and enable
the exploration of combinatorially growing compositional spaces with ease.

\subsection{Diatomic series with 14 electrons}

\begin{figure}
\includegraphics[width=8.5cm]{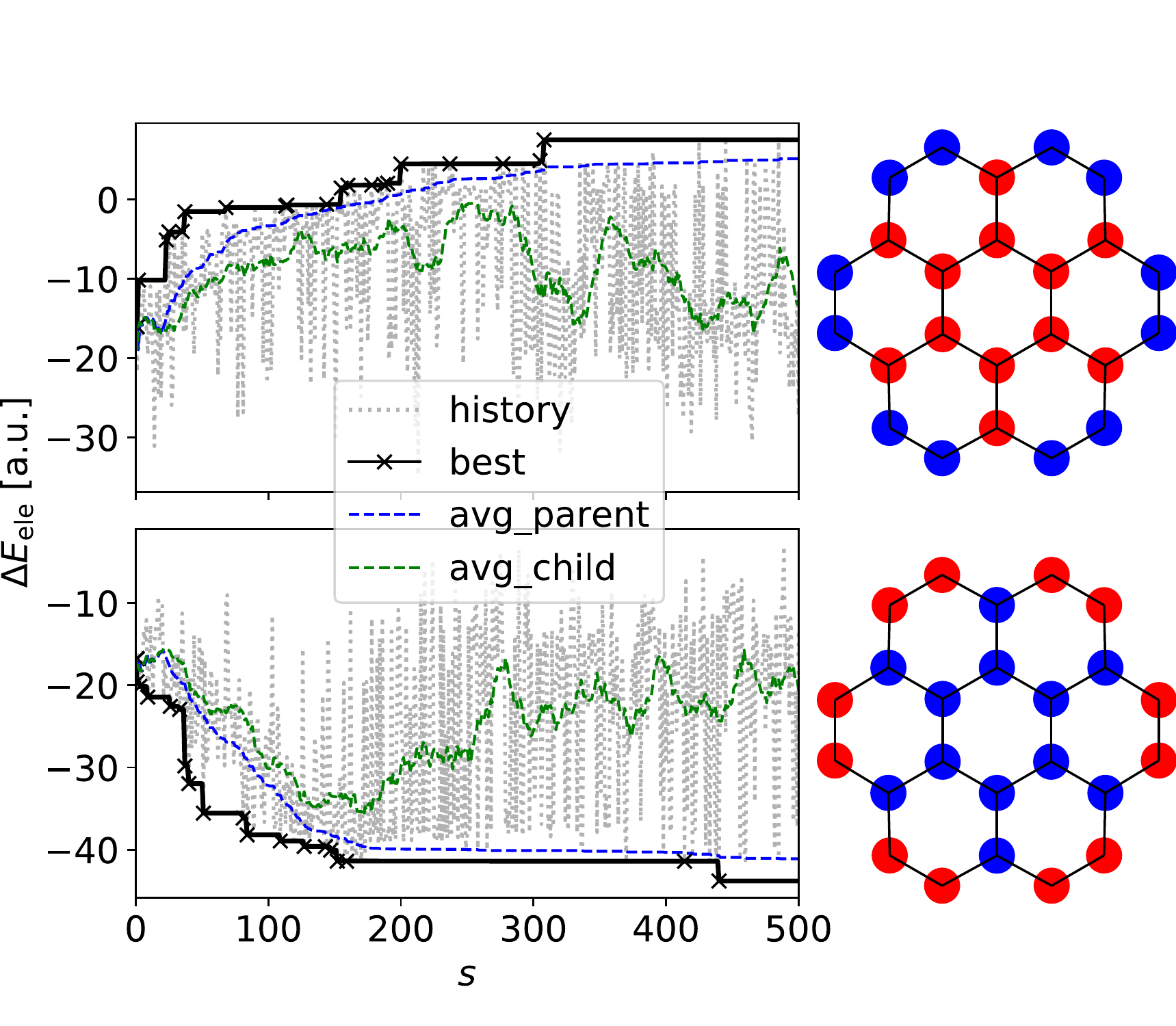}
\caption{Genetic algorithm maximization (upper panel) and minimization (lower panel) on electronic energy for (BN)$_{12}$H$_{12}$. 
The difference electronic energy for
optimization history (gray dotted lines), current optimal molecule (black crosses), average over the parent pool (blue dashed lines) and the corresponding children (green dashed lines) are plotted respectively.
The optimized molecules are shown on the right where B and N are represented by red and blue atoms.}
\label{fig:Opt1}
\end{figure}

\begin{figure}
\centering
\includegraphics[width=8.5cm]{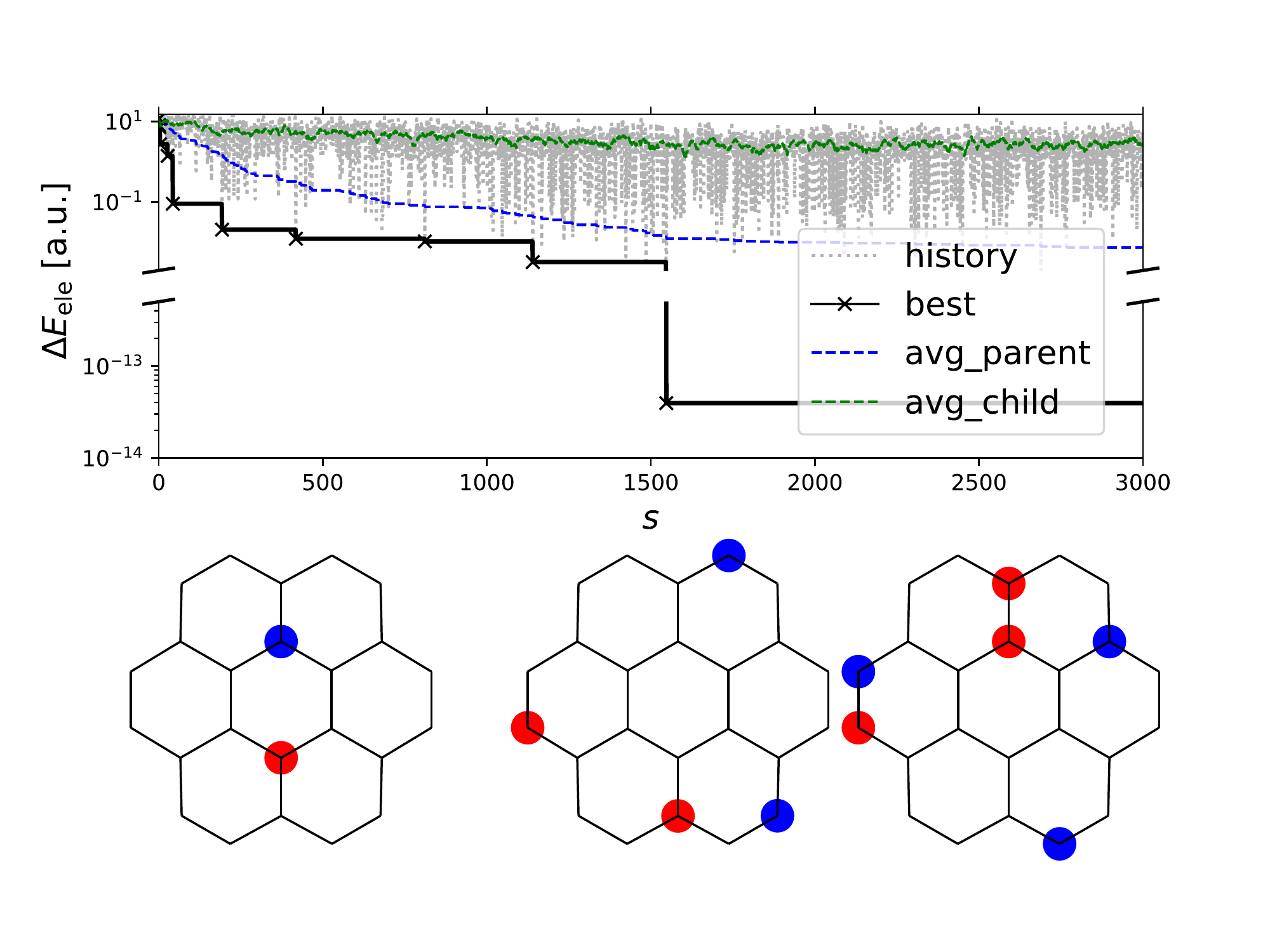}
\caption{
Genetic algorithm energy optimization run. The target energy corresponds to left hand molecule.
The black line (best) corresponds to the best molecule in population. The blue line (average)
corresponds to the average value of the population.
The target molecule is discovered within 7 steps out of $\sim$4 billion compounds. 
Energies of mid and right hand side molecule deviate only by 0.4 and 0.5 mHa from target.
}
\label{fig:Opt2}
\end{figure}

The projection of the unified ANMs based structure of CCS onto lower dimensional realistic 
systems is straight-forward. We exemplify this for the electronic energy of the
neutral iso-electronic $N =$ 14 electron series with 
variable $\fatZ$ and $\{\fatR_I\}$, as shown in Fig.~\ref{fig:N2}.
Note the text-book dependence of the electronic energy on interatomic distance
(decaying towards the united atom energy of Si),
the well known concavity for fixed $N$ and $\{\fatR\}$~\cite{anatole-ijqc2013}, 
and the ridge corresponding to the homo-diatomic N$_2$ ($\Delta Z$ = 0). 
The apparent monotonic and smooth behavior of the electronic energy in this sub-space 
corroborates the applicability of perturbation theory. 
This implies that, in analogy to vibrational normal modes, 
the gradient with respect to $Z$ must be zero at the ridge. 
And it is obvious, indeed, that the corresponding Hellmann-Feynman derivative, 
$\int d\fatr \, \rho(\fatr) (\frac{1}{|\fatr-\fatR_1}-\frac{1}{|\fatr-\fatR_2|})$~\cite{anatole-jcp2009-2}, 
must be zero due to the symmetry of the electron density, just as well as
all higher odd order energy derivatives.
As such, when mutating nuclear charges, the reference system with symmetrical atomic densities
will always correspond to a maximum in the electronic energy $E$. 
This observation would suggest that it is preferable to select 
reference systems with maximal symmetry in order to quench odd higher order effects. 

Numerical electronic energy estimates of alchemically adjacent systems $|d\fatZ| = 2$
within the same neutral iso-electronic 
diatomic series with $N =$ 14 electrons (lower panel in Fig.~\ref{fig:N2})
support this idea: The prediction error increases systematically 
for estimates of CO, BF, and BeNe when using systems as reference which
decrease in symmetry, i.e.~N$_2$, CO, and BF, respectively.
These results confirm, not surprisingly, that the harmonic approximation 
works best at the ridge, in complete analogy to harmonic 
vibrational normal-modes working best at zero Kelvin.
The error becomes largest for changes involving substantial changes in
electron densities, e.g.~when valence electrons flow from $p$ to $s$ orbitals and from principal
quantum number 2 to 1 and 3 (for example when targeting or referencing BeNe, LiNa, or HeMg).
Also note the negative sign of the error as one predicts CO from N$_2$, BF from CO, and BeNe from BF.
This implies an exponent of the actual energy surface which is larger than 2, which is in line
with independent findings for the energy of free atoms scaling as $\sim -Z^{7/3}$~\footnote{K.~Burke, oral contribution, IPAM reunion 2018}. 
It is also interesting to note the left/right anti-symmetry in the error 
of most predictions, e.g.~the error made when predicting the electronic energy of CO
using BF as a reference has the same magnitude as its reverse counterpart, 
i.e.~predicting the electronic energy of BF using CO as a reference.
This suggests, that the exponent is not much larger than 2. 
And it is to be contrasted with the findings for first order based estimates of energy changes,
e.g.~for alchemical predictions of covalent bond energies~\cite{Samuel-CHIMIA2014,Samuel-JCP2016},
where, due to the concavity of the electronic energy in $Z$, 
the error is clearly not symmetrical upon exchange of reference and target system.



\subsection{BN doping of benzene}

While any iso-electronic diatomic series can be expanded in the ANMs of the corresponding homo-diatomic, 
the maximum ridge in the electronic energy, ANMs become less obvious for larger molecules. 
Based on above symmetry arguments, the benzene molecule with point group
$D_{6h}$ emerges as an intuitive 2D poly-atomic reference system. 
Considering all the possible neutral iso-electronic changes of carbon to B and N
it is clear that some odd order energy derivatives will be zero due to symmetry. 
Scaling up the coordinates will lead to the electronic energy of the dissociated free atoms, 
while scaling them down leads to nuclear fusion, i.e.~the energy of the united atom, Mo ($Z =$ 42).
Here, we remind the reader that we do not consider nuclear Coulomb repulsion, 
and that for this and the remaining examples, ANM based predictions are
always exemplified for changes in nuclear charges only, i.e.~keeping coordinates and electron numbers fixed.

Fig.~\ref{fig:benzeneAll} illustrates eigenvalues and ANMs of benzene and their use for predicting the iso-electronic doping of benzene with B and N 
(keeping number of electrons and geometry constant) at all possible atomic sites, 
i.e.~for all possible constitutional isomers with sum formula C$_4$BNH$_6$, C$_2$(BN)$_2$H$_6$, and (BN)$_3$H$_6$.
It is intriguing to note the similarity of form, degeneracy, and energy ordering to ordinary H\"uckel orbitals of benzene. 
Also note that, in analogy to H\"uckel, the eigenvalues decrease as the number of nodes in the ANM decrease.
Obviously, however, there is no $\pi$-electron structure at the atom's origin,
and also the eigenvalues do not correspond to solutions to H\"uckel's secular equation. 
While the eigenvalues depend on the level of theory used, because of symmetry the ANMs are independent of that. 

Apart from their appealingly simple and insightful structure, 
one can use these ANMs to easily estimate relative energetics of possible mutations on the back of an envelope. 
For example, using ANM based CCSD predictions of the electronic energies according to Eq.~\ref{eq:Predict}, 
i.e.~$E(\fatx^t) \approx \sum_i \epsilon_i c_i^2$, of the three constitutional isomers of BN 
doped benzene result in -1.226, -1.312, -1.348 Ha for ortho, meta, and para substitutions, respectively. 
Compared to actual values, the energetic ordering is conserved, and the estimates are in decent agreement 
with the corresponding CCSD energies (-1.275, -1.362, -1.401 Ha), i.e.~systematically overestimating the truth by $\sim$0.05 Ha.
Predicted and actual changes in energy with respect to pure benzene are also
on display in Fig.~\ref{fig:benzeneAll} for all the possible mutants, and 
indicate very decent qualitative agreement.
Qualitatively, the energetic order can also be explained by noting that 
the closer the poles of the perturbing potential, 
the smaller the integral of their product with the electron density response, 
the smaller the deviation from the energy of benzene.
This is consistent with the fact that ANMs with fewer nodes have lower eigenvalues.
By consequence, the isomers with sum formula (BN)$_3$H$_6$ will decrease in
energy when decreasing the number of nodal surfaces between B and N mutations, i.e.~
$E$(B$_3$N$_3$H$_6$) 
$<$ $E$(B$_2$NBN$_2$H$_6$) 
$<$ $E$(BNBNBNH$_6$).
Inspection of the linear combination of ANMs resulting in each of these
isomers also clearly indicates that the energy decays as ANMs 
with fewer nodes are being blended in.
Unfortunately, before one can compare to experiments, 
addition of the nuclear repulsion terms will obfuscate this ranking unless
the inequalities introduced by Mezey can be applied~\cite{ConcavityMezey1985}.
We believe nevertheless that these rules are obviously useful for estimating the ranking 
of electronic energies in constitutional isomers which is of utmost relevance
for gaining a deepened and more intuitive grasp of quantum chemistry based relationships. 
We do not think that these rules have been noted yet.

\begin{figure}
\includegraphics[width=8.5cm]{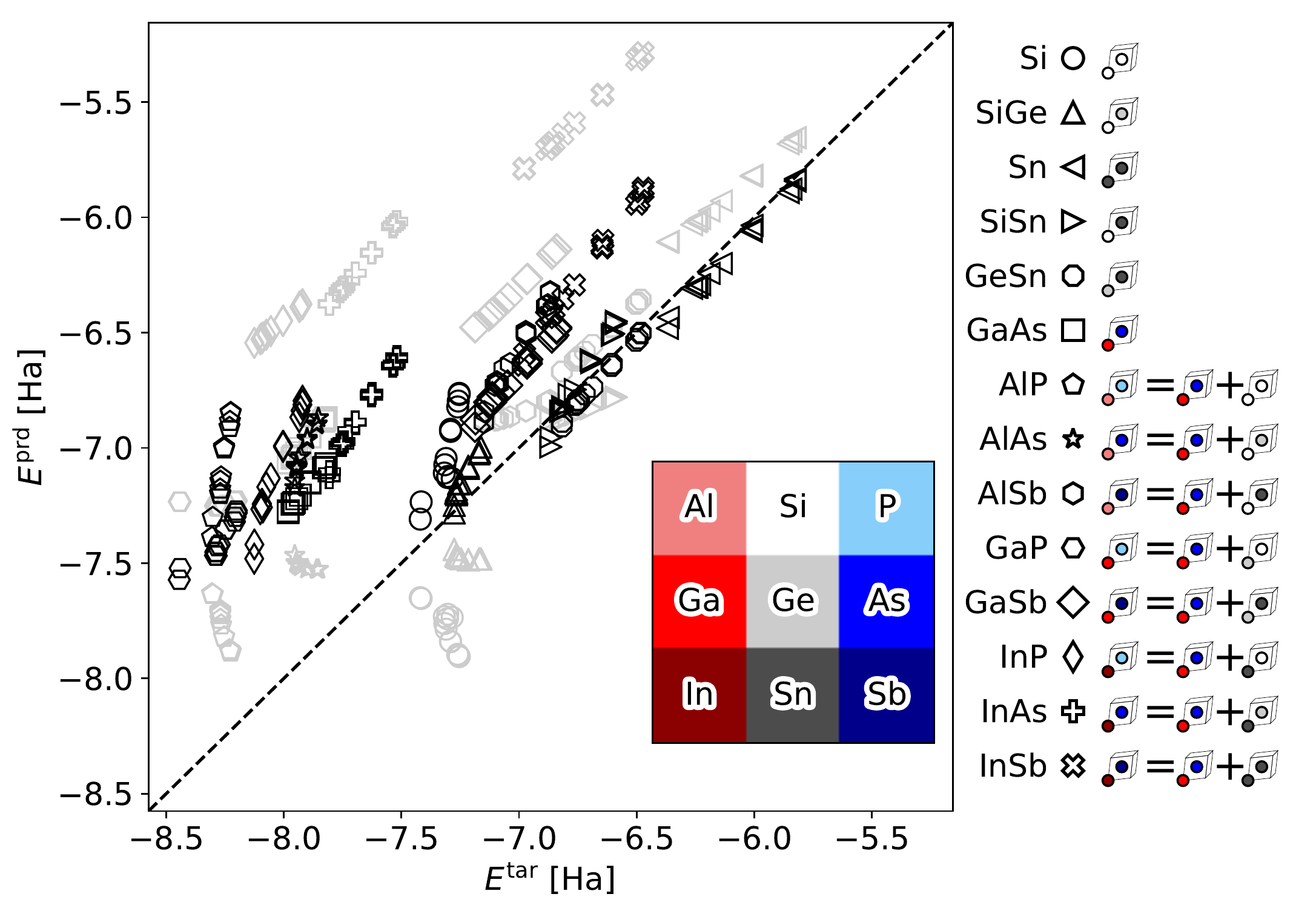}
\caption{
Alchemically predicted vs.~true scatter plot 
for first (gray) and second order (black) estimates of total energies for various III-V and IV-IV 
semiconductors expanded in Ge ANMs in periodic table (blue/red within period, 
white/black within column).
}
\label{fig:Ge}
\end{figure}

\subsection{BN doping of coronene}

In order to explore the applicability of this approach to the computational design problem
of real materials we have considered the case of BN doping also for 
coronene (C$_{24}$H$_{12}$), relevant for molecular electronics applications~\cite{Andrienko2008,AndrienkoKremerMullen2009}.
BN doping of coronene results in the 24 ANMs and eigenvalues shown in the upper
panel of Fig.~\ref{fig:Diagram2}. 
Coronene possesses three symmetrically distinct classes of carbon atoms:
Six atoms corresponding to the inner ring, six atoms bonded to the inner ring, and the twelve outer atoms. 
Consequently, first order derivatives with respect to iso-electronic BN doping within any of these
three groups of atoms are zero.
Doping with BN pairs within these classes, we have generated over 1.8 M mutants, 
and estimated their energy based on coronene's ANMs. 
Duplicates have been removed using the sorted Coulomb matrix representation~\cite{RuppPRL2012,AssessmentMLJCTC2013}.
The resulting energies are shown in ascending order in the lower panel of Fig.~\ref{fig:Diagram2},
together with a sub-sample of more than two thousand validating cases for which the corresponding DFT PBE 
energy has been calculated.
Clearly, the overall qualitative trends of DFT and ANM based estimates agree well with each other. 
A linear fit of predictions to validations for the 2 k mutants yields a MAE of $\sim$28 kcal/mol.
Further analysis indicates that the error grows with number of BN pairs 
by, on average, $\sim$6.5 kcal/mol per BN pair.
In the case of the hundred coronene mutants closest in energy to coronene, 
for example, the MAE amounts to only $\sim$2.1 kcal/mol.
Fig.~\ref{fig:Diagram2} also reports the energy distribution of the 1.8 M coronene mutants. 
Discrete peaks correspond to higher lying mutant stoichiometries with less BN content.

We have also explored the usefulness of ANM based energy estimates 
for the molecular design challenge of finding those constitutional isomers
of (BN)$_{12}$H$_{12}$ with the respectively lowest and highest electronic energy.
A genetic optimization algorithm based on first (which now can be non-zero
due to BN doping among symmetrically inequivalent carbon atoms)
and second order energy estimates only converges within a few hundred steps.
The optimization history, together with the converged molecules, are on display in Fig.~\ref{fig:Opt1}.
The most and least stable isomers correspond, not surprisingly, to those N and B distributions
which localize the valence electron density most and least, respectively.
Results in Fig.~\ref{fig:Opt2} summarize the genetic optimization history when searching for the
energy of the coronene mutant resulting from BN doping in para position of the inner carbon ring. 
The target molecule is identified by the genetic optimizer after just over 1500 optimization steps.

These calculations can serve to illustrate the scope of the computational savings which 
result from the use of ANMs: $\sim$1000 CPU core hours were necessary on average to calculate
the over 2 k validating DFT energies. The ANM based estimates of 1.8 M mutants, 
by comparison, incurred negligible overhead ($\sim$30 CPU core minutes).

\subsection{Expanding III-V and IV-IV semiconductors in ANMs of Ge}

Finally, we have investigated the applicability of ANMs to solids. 
More specifically, we have considered iso-valence-electronic expansions in the ANMs resulting 
from a minimal unit cell in the pseudopotential parameter space of two Ge atoms.
Using 15 parameters in the analytical pseudopotentials of Goedecker and Hutter~\cite{SG,krackPP},
four dimensions have been considered per atom: Right and left in a period of the periodic table 
(analogous to $Z$ as discussed above), and up and down in a column of the periodic table 
(corresponding to changes in principal quantum number). 
The pseudopotential parameters were coupled to these dimensions with the chain-rule,
as explained in the Methods section. 
The resulting projection yields the ANMs of Ge used for expansion as shown
in Fig.~\ref{fig:Ge} for all possible IV-IV and III-V semiconductors which neighbor Ge 
in the periodic table, i.e.~Si, Sn, SiGe, SnGe, SiSn, as well as AlP, AlAs, AlSb, GaP, 
GaAs, GaSb, InP, InAs, and InSb.
More specifically, the figure shows first and second order total potential energy 
estimates vs.~actual DFT evaluations for various lattice scans. 
Note that first order gradients are non-zero due to lack of symmetry in 
pseudopotential parameter space, i.e.~while the reference electron density is 
symmetric, the perturbing potential is not perfectly anti-symmetric.
A clear correlation is found for first order estimates. 
Inclusion of second order contributions through ANM based predictions
improves the overall correlation, and, maybe more importantly, 
results in a systematic overestimation of the energies of the 
target systems (consistent with aforementioned observations made for molecules).
The prediction quality for IV-IV crystals is particularly encouraging, 
in all likelihood profiting from near-linear changes in valence electron density as
one changes from one period to the next. 
ANM based estimates of III-V materials, however, are more challenging. However, the
errors appear to be rather systematic in their overestimation. This raises hope that it can still 
be quite feasible to correct it once third order contributions are being included.

\section{Conclusions}
The Born-Oppenheimer approximation implies a parametric dependency of the electronic
ground-state energy on nuclear positions, charges, and electron number. 
In order to obtain a general yet rigorous framework of chemical space, we have unified 
all the relevant degrees of freedom by extending the ordinary normal mode procedure used for atomic positions 
by alchemical normal modes (ANMs) which also include nuclear charges and electron number. 
Applied within Taylor expansions, the energy of iso-electronic target compounds can be 
expanded in ANMs. The resulting estimates are exact up to third order 
if electron densities in reference system and perturbing potential are symmetric and anti-symmetric, respectively. 
We have illustrated the concept for diatomics using molecular nitrogen as a reference,
for all and $\sim$1.8 M BN doped mutants of benzene and coronene, respectively. 
The applicability to solids has been demonstrated for all III-V and IV-IV
semiconductors neighboring GeGe.
Future extension to higher orders to improve predictive power,
to other properties, excited states, and
generalizations to entire functional groups can also be envisioned.

\section{Methods}
\label{sec:methods}
\section{Unified Hessian matrix for N$_2$}

The full Hessian matrix as defined in Eq.~\ref{eq:full_hessian} for N$_2$ at 
its equilibrium geometry 
(using PBE~\cite{PBE}  with uncontracted cc-pVDZ basis function in Gaussian09~\cite{Gaussian09}).
All matrix elements are rounded to the third decimal numbers.

\begin{equation}
\begin{array}{c}
\\[2pt]\mathbf{H} = \ \
\end{array}
\begin{blockarray}{ccccc}
\Delta Z_1 & \Delta Z_2 & \Delta \mathbf{R} & \Delta N \\
\begin{block}{(cccc)c}
-3.126 & 0.139 & 0.121 &-0.575 & \Delta Z_1\\
 0.139 &-3.126 & 0.121 &-0.575 & \Delta Z_2\\
 0.121 & 0.121 &-9.477 &-0.121 & \Delta \mathbf{R}\\
-0.575 &-0.575 &-0.121 & 0.139 & \Delta N\\
\end{block}
\end{blockarray}
\end{equation}

The corresponding eigenvalues and eigenvectors are

\begin{equation}
\begin{blockarray}{ccccc}
\mbox{ANM}_1 & \mbox{ANM}_2 & \mbox{ANM}_3 & \mbox{ANM}_4  \\
\begin{block}{(cccc)c}
-0.018 & 0.168 & 0.707 & 0.687 & \Delta Z_1 \\
-0.018 & 0.168 &-0.707 & 0.687 & \Delta Z_2 \\
 1.    & 0.016 &-0.    & 0.022 & \Delta \mathbf{R} \\
 0.01  &-0.971 &-0.    & 0.238 & \Delta N \\
\end{block}
-9.48 & 0.34 & -3.27 &-3.18 & \epsilon \\
\end{blockarray}
\end{equation}

where ANM$_1$ and ANM$_2$ are mostly changing $\mathbf{R}$ and $N$ with eigenvalues -9.48 [Ha/bohr$^2$] and 0.34 [Ha/$e^2$] respectively. ANM$_3$ (eigenvalue -3.27 [Ha/$e^2$]) is purely antisymmetric in changes in $Z_1$ and $Z_2$.  ANM$_4$ (eigenvalue -3.18 [Ha/$e^2$]) is remarkable: one can see how in this principle component the increase of the nuclear charge on both atoms requires also an increase in number of electrons (to compensate the change in Z) {\em and} requires a slight increase of the distance between the atoms.
It should be noted that due the seminegative definite nature of the linear response function, the eigenvalues with respect to the changes in $\mathbf{R}$, $Z$ are all negative.

\subsection{Computational details}
For the diatomics and coronene, we used PBE~\cite{PBE}  with uncontracted cc-pVDZ Ne basis function for all atoms in Gaussian09~\cite{Gaussian09} and
HORTON~\cite{HORTON} for all molecular examples.

For benzene, we used CCSD/cc-pvdz in Gaussian09~\cite{Gaussian09}
with the massage keyword to modify the nuclear charges. As such, the Carbon cc-pvdz basis functions were also used for Nitrogen and Boron (uncontracting it was computationally very expensive and created instabilities we could not resolve). To calculate the curvature, Benzene was calculated +\/- 0.25$\times$eigen vector. So for the completely symmetric A1g, this means the carbon nuclear charge +0.25$\times 1/\sqrt{6}$ = 6.1020620725.

For the energy calculations of solids, a 1x1x1 face-centered cubic (fcc) primitive super cell with two atoms was used (no k-point sampling)
within the plane-wave basis set code CPMD~\cite{cpmd3.15}, in combination with 
the PBE~\cite{PBE} functional, a plane-wave cutoff of 100 Ha,
and Goedecker-Teter-Hutter pseudopotentials~\cite{SG,krackPP}.

\subsection{ANM based estimates}
One can expand the potential energy ground state hyper surface of any target system $E_t(\{\fatR,Z\})$ around a 
symmetric iso-electronic reference system with identical atomic coordinates and 
energy $E_0$ by means of a Taylor expansion in coupling parameter $0 \le \lambda \le 1$,
\bea
E_t(\{\fatR,Z\}) &\approx & E_0 + \frac{1}{2} \frac{\partial^2 E_0}{\partial \lambda^2} d\lambda^2 + {\rm EHOT} \\
E_t(\{\fatR,Z\}) &\approx & E_0 + \frac{1}{2} \sum_{IJ}\frac{\partial^2 E_0}{\partial Z_I \partial Z_J} dZ_J dZ_I \nonumber\\  &&+ {\rm EHOT} \nonumber \\
\label{eq:expand}
\eea
for $\partial_\lambda Z_J = dZ_J$ and for $d\lambda = 1$,
and EHOT corresponding to even higher order terms.
Note that instead of nuclear charges $\{Z_I\}$ 
one can use pseudopotential parameters $\{\sigma_i\}$ just as well.
To simplify this equation, let $\mathbf{Q}$ be the matrix of the eigenvectors of the second order derivative matrix $\mathbf{H}_{IJ}=\frac{\partial^2 E_0}{\partial Z_I\partial Z_J}$ (or Hessian),
\beq
\mathbf{HQ} = \mathbf{Q \Upsilon}
\eeq
where $\mathbf{\Upsilon}$ is a diagonal matrix with the eigenvalues $\epsilon_m$ of $\mathbf{H}$. The eigenvalues $\epsilon_m$ are solutions of the alchemical secular equation,
\beq
{\rm det}(\partial^2_{Z_I,Z_J} E_0 - \delta_{IJ}\epsilon_m) = 0
\eeq

We now define the alchemical normal mode vector $Q_i$ consisting of the columns of $\mathbf{Q}$. 
For a given target molecule, changes in the nuclear charge vector ($d{\bf z} = \partial_\lambda {\bf z}(\lambda)$), 
can be expressed in the new basis of the alchemical normal modes as a linear combination:
\bea\label{eq:c=Qz}
{\bf c} &=& {\bf Q} d{\bf z}
\eea
resulting in 
\bea\label{eq:d2U=sum_ec2}
\sum_{IJ}\frac{\partial E_0}{\partial Z_I \partial Z_J} dZ_I dZ_J & = & \sum_i \epsilon_i c^2_i
\eea

The energy along a chosen alchemical path $\mathbf{z}(\lambda)$ can be expressed as $E(\mathbf{z}(\lambda))$ where the corresponding alchemical derivative at the reference system $0$ is 
\beq
\partial_\lambda U_0\big(\mathbf{z}(\lambda)\big) = \nabla E_0(\mathbf{z})\cdot d\mathbf{z} = \sum_I (\partial_{Z_I}E_0)\partial_\lambda Z_I,
\eeq
where $E_0(\mathbf{z}) = E_0(Z_1, \cdots, Z_N)$ is a $\mathbf{R}^N\mapsto\mathbf{R}$ function described by $\{Z_I\}$.

Within the orthogonalization transformation, the basis is changed from nuclear charges $\{Z_I\}$ to alchemical normal modes $\{Q_i\}$ where the magnitude in each dimension $c_i$ denotes the amplitude of each normal mode. In other words, the energy is rewritten as
\beq
E_0(\mathbf{z}) \Rightarrow E_0(\mathbf{c}) = E_0(c_1, \cdots, c_N).
\eeq
Notice that $\mathbf{c}$ is a linear function in $\lambda$ due to Eq.~(\ref{eq:c=Qz}) where $\mathbf{Q}$ is independent of $\lambda$ and $\mathbf{z}$ is linear in $\lambda$.

The alchemical derivative within alchemical normal mode basis is 
\beq
\partial_\lambda E_0\big(\mathbf{c}(\lambda)\big) = \sum_i (\partial_{c_i}E_0)\partial_\lambda c_i.
\eeq
Notice that $\partial_\lambda c_i = \sum_J Q_{iJ}\partial_\lambda Z_J$.
And the second order derivative is
\beq
\begin{array}{rcl}
\partial_\lambda^2 E_0 
&=&\displaystyle 
\partial_\lambda\Big(\sum_i \frac{\partial E_0}{\partial c_i}\frac{\partial c_i}{\partial\lambda}\Big)\\
&=&\displaystyle 
\sum_i\Big(\sum_j\frac{\partial}{\partial c_j}\frac{\partial E_0}{\partial c_i}\frac{\partial c_j}{\partial\lambda}\Big)\frac{\partial c_i}{\partial\lambda}\\
&=&\displaystyle 
\sum_{ij}\frac{\partial^2 E_0}{\partial c_i\partial c_j}(\partial_\lambda c_i)(\partial_\lambda c_j).
\end{array}
\eeq

Notice that $\partial_\lambda^2 c_i = 0$ because $d\mathbf{z}$ is linear in $\lambda$.
And $\frac{\partial^2E_0}{\partial c_i\partial c_j} = \delta_{ij}\epsilon_i$ is the diagonal matrix and it is connected to $\frac{\partial^2E_0}{\partial Z_I\partial Z_J}$ via
\beq
\Big(\frac{\partial^2E_0}{\partial c_i\partial c_j}\Big)_{ij} = \mathbf{Q}^T\Big(\frac{\partial^2 E_0}{\partial Z_I\partial Z_J}\Big)_{IJ}\mathbf{Q}
\eeq

And the Eq.~(\ref{eq:d2U=sum_ec2}) can be rewritten as
\beq
\begin{array}{rcl}
\displaystyle
\sum_{IJ}\frac{\partial E_0}{\partial Z_I \partial Z_J}dZ_I dZ_J
&=&\displaystyle
d\mathbf{z}^T\Big(\frac{\partial^2E_0}{\partial Z_I\partial Z_J}\Big)_{IJ}d\mathbf{z}\\
&=&\displaystyle
d\mathbf{z}^T\mathbf{Q}^T\Big(\frac{\partial^2E_0}{\partial c_i\partial c_j}\Big)_{ij}\mathbf{Q}d\mathbf{z}\\
&=&\displaystyle
\mathbf{c}^T
\Big(\frac{\partial^2E_0}{\partial c_i\partial c_j}\Big)_{ij}
\mathbf{c}\\
&=&\displaystyle \sum_i \epsilon_i c_i^2
\end{array}
\eeq

\subsection{Pseudopotential space}
When pseudopotentials (PP) are used,
ANM space is spanned by the PP parameters. 
For fcc primitive cell of two atoms, there are 30 parameters (15 per atoms).
The Hessian matrix elements can be approximate by finite difference

\beq
\begin{array}{rcl}
\mathbf{H}_{ij}
&=&\displaystyle
\frac{\partial^2 E}{\partial\sigma_i\partial\sigma_j}
\\&=&\displaystyle
\frac{\partial}{\partial\sigma_i}\Big(\frac{\partial E}{\partial\sigma_j}\Big)
\\&\approx&\displaystyle
\frac{\partial}{\partial\sigma_i}\Big(\frac{E(\sigma_i, \sigma_j+\Delta\sigma_j) - E(\sigma_i, \sigma_j)}{\Delta\sigma_j}\Big)
\\&\approx&\displaystyle
\frac{1}{\Delta\sigma_i}
\Bigg(\frac{E(\sigma_i+\Delta\sigma_i, \sigma_j+\Delta\sigma_j) - E(\sigma_i+\Delta\sigma_i, \sigma_j)}{\Delta\sigma_j}
\\&&\displaystyle
- 
\frac{E(\sigma_i, \sigma_j+\Delta\sigma_j) - E(\sigma_i, \sigma_j)}{\Delta\sigma_j}\Bigg).
\end{array}
\eeq

That is, there are four finite difference calculations required for each of the matrix elements:
$E(\sigma_i+\Delta\sigma_i, \sigma_j+\Delta\sigma_j)$, 
$E(\sigma_i+\Delta\sigma_i, \sigma_j)$, 
$E(\sigma_i, \sigma_j+\Delta\sigma_j)$, 
$E(\sigma_i, \sigma_j)$, 
where only the first term is unique for each element.

Note that the finite difference formula is different for diagonal terms $\mathbf{H}_{ii} = \frac{E(\sigma_i+\Delta\sigma_i) - 2E(\sigma_i) + E(\sigma_i-\Delta\sigma_i)}{\Delta\sigma_i^2}$.
The required finite difference calculations are
\begin{itemize}
\item $E(\sigma_i+\Delta\sigma_i, \sigma_j+\Delta\sigma_j)$: $N(N-1)/2$ calculations for $i\neq j$.
\item $E(\sigma_i+\Delta\sigma_i)$: $N$ calculations.
\item $E(\sigma_i-\Delta\sigma_i)$: $N$ calculations.
\item $E(\sigma_i, \sigma_j)$: 1 calculation
\end{itemize}

which adds up to $\frac{N^2}{2}+\frac{3}{2}N+1$ calculations, where $N$ is the number of parameters in the system.


\section{Acknowledgements}
We would like to thank F.~A.~Faber and G. F. von Rudorff for insightful discussions. 
OAvL acknowledges support by the Swiss National Science foundation 
(No.~PP00P2\_138932, 407540\_167186 NFP 75 Big Data, 200021\_175747, NCCR MARVEL).
Some calculations were performed at sciCORE (http://scicore.unibas.ch/) 
scientific computing core facility at University of Basel.

\bibliography{literatur}{}
\bibliographystyle{ieeetr}

\end{document}